# RADIATIVE PHASE TRANSITIONS AND THEIR POSSIBLE ROLE IN BALANCE OF ATMOSPHERE HEATING

Mark E. Perel'man [1)]

*Racah Institute of Physics, Hebrew University, Jerusalem, Israel*

**ABSTRACT**. Condensation and sublimation of water vapors (and $CO_2$, $CH_4$, $N_2O$ vapors also) in the Earth atmosphere must be accompanied by emission of latent heats on characteristic frequencies marked in absorption spectra. Calculated wave lengths completely explain all peaks observed for these gases in the near IR. Established phenomena require further investigations, re-estimation of atmospheric heat balances and so on. Investigation of analogical peaks in atmospheres of other planets can be used for analyses of their structures.

---------------------

## INTRODUCTION

At consideration and comparison of different channels of heat removing from the Earth is usually assumed that latent heat of precipitations (the power of 78 W/cm$^2$ from 324 W/cm$^2$ of general flux) is accumulating directly directly on the Earth surface (e.g. [1]).

Let us show that possible processes of repeated emissions of latent heats at condensation and crystallization of certain atmospheric gases with their subsequent sublimation and anew solidifications can lead to augmentation of the role of this channel of the heat removal from the Earth.

Can go this removing in the form of usual black body radiation or a more characteristic special possibility may exist? Is all latent heat directly steaming out on the surface of Earth or, at least partly, in the atmosphere on higher altitudes? If it is executed on characteristic frequencies, it can carry away by emission through atmosphere more energy than is usually supposed.

Processes of condensation (solidification, crystallization) and processes of sublimation (boiling, melting) are interconnected by the thermodynamics balance relation. However, at the microscopic approach the basic difference between direct and opposite processes should be taken into account. So, the removal of particle from condensate demands a concentration of energy on its bonds which at $T < T_c$ is much greater the average thermal energy, therefore such process must be connected to accumulation, possibly gradual, of fluctuations of the thermal spectrum.

[1]). E-mail: m.e.perelman@gmail.com

But the connection of particle with the condensate, i.e. the entering of particle into it, demands a removal of bonds energy by electronic transitions, i.e. an emitting of quanta with energy greater average energy of thermal spectrum which, the truth, can be further thermalized.

In the articles [2, 3] was theoretically proved that latent energies at the phase transition of the first kind can or even must be emitted on the characteristic frequencies determined via latent energy per atom/molecule (by new forming bonds). This approach is based on the consideration of different phases in the frame of quantum electrodynamics (QED) as series of electron levels, transitions between them would be executed by photons, real or virtual, and therefore the removing of latent energy at transitions onto down levels must be naturally executed by photons emission with their possible further thermalization. In the more general sense this theory leads to description of interactions of all constituents in substance via photons exchange and to the first adequate determination of critical indices, etc. [4].

In accordance with the microscopic point of view, phase transitions in more ordered states can be alternatively described as a sequence of entering of particles into new phase. As each particle can be characterized by the natural or inducing electromagnetic moment, such transition into medium with another, generally speaking, electromagnetic susceptibility, should be accompanied by transitive radiation [3] (some details of theory with an overview of experimental data are adducing in [5]). Both approaches lead to the identical results.

The theory was experimentally verified at condensation of water vapor [6] and aluminum and copper vapors [7], apart from them there are certain observations that can be interpreted as the radiation of latent heat at water vapor condensation in atmosphere [8] and in laboratory [9]. The idea of characteristic emission of latent heat at gases condensation leads to the explanations of the great IR excess in spectra of young nebulae and stars [3]. More experimental data was received with examination of crystallization processes (they are overviewed in [5]).

Note that at passing through non-transparent substances these characteristic frequencies can be thermalized, i.e. can be transformable into phonons and so on that evidently complicates laboratory investigations. But in the atmosphere a part of them, in an ideal case half of them, at sublimation within upper layer of clouds, can be transferred via transparent higher atmosphere and removed from the Earth. Therefore anyway it must be taken into account for examination of the atmospheres heat balance.

**SPECTRA OF EMISSION AT PHASE TRANSITIONS**

Let's estimate of possible energy of one photon emitted by single particle at its entering into condensate, i.e. at the elementary act of phase transition (temperature remains constant):

$$\hbar\omega_1 = \Lambda/N_A, \qquad (1)$$

where $\Lambda$ is the molar latent energy, $N_A$ is the Avogadro number. (We speak about emission only, but all described frequencies should be marked naturally in absorption spectra.)

But here must be taken into account that really the emission is connected with formation of bonds between an entering particle and particles in the condensate; for comparatively simple particles can be assumed that all such bonds are identical. As the molecules $H_2O$ are connected with neighbors by four bonds, at water condensation each molecule establishes two new bonds with the energy $\sim \hbar\omega_1/2$. But as in the process of emission energies of some bonds can be accumulated and then emitted as one quantum and such joining can include dimers and even more complex formations, it can be proposed that quanta with energies

$$\hbar\omega_n \sim n\Lambda/N_A \qquad (2)$$

would be observable, where *n* is the number of formatting bonds of single molecule or their aggregate. This expression can be rewritten in a more usable form via the wave length:

$$\lambda_n \sim 120/n\Lambda, \qquad (3)$$

where $\Lambda$ is expressed in kJ/mole, and lengths of waves are in mcm.

All observable peaks are very wide and this feature must be discussed. At laboratory investigations the variation of energies of condensable or crystallizing particles are restrained by usual equilibrium distributions. But in the case of atmospheric observations there is a natural distribution over altitudes and temperatures. For initial estimation the famous Trouton rule can be taken into account: the temperature and the mean wave-line of thermal emission at the gas condensation are connected by the relation $\lambda T \approx 0.29$ cm K (its theoretical deduction is given in [4, 5]). It can be supposed that analogical relation, with another and now still undetermined numerical value can be proposed for transitive emission also. Therefore for considered cases can be assumed such estimation:

$$\Delta\lambda/\lambda \sim \Delta T/T. \qquad (4)$$

Differences of temperatures within atmosphere evidently can lead to observable values of peaks wideness.

**THE EARTH ATMOSPHERE**

Let's compare experimental (observable, e.g. [1]) data with estimations by the general expression (3) for water vapor.

In the Table are written out all its observable maxima from 1 till 6 mcm; they are compared with wave lengths calculated by (3) at $n = 1 \div 7$ with the heat of sublimation $\Lambda^{(subl)} = 46.68$ kJ/mole and the heat of condensation into liquid phase at normal conditions $\Lambda^{(cond)} = 40.6$ kJ/mole (the heat of water crystallization $\Lambda^{(cryst)} = 6.01$ kJ/mole leads to a far IR and is omitted).

**Table**

| $\lambda_{observ}$ | 0.72 0.81 | 0.93 | 1.13 | 1.42 | 1.89  2.01-2.05 | 2.25-3.0 | 5.9 |
|---|---|---|---|---|---|---|---|
| n | 7 | 6 | 5 | 4 | 3 | 2 | 1 |
| $\lambda_n^{(subl)}$ | 0.73 | 0.86 | 1.03 | 1.285 | 1.71 | 2.57 | 5.14 |
| $\lambda_n^{(cond)}$ | 0.84 | 0.985 | 1.18 | 1.48 | 1.97 | 2.96 | 5.91 |

All these lines are between radiated wave lengths corresponding to mixture of condensing and sublimation processes. The radiation with $\lambda \sim 2$ mcm corresponds to dimers or even more complex formations (notice that by their intensities relative to density of dimers can be estimated and so on). Note evident possibilities of splitting these maxi's onto condensation and sublimation types.

In this range are another max's also. The peaks with λ = 3.1 mcm and with λ > 6.5 mcm, that are not written out here, correspond to the well-known proper vibrations of molecule $v_1$ (λ = 3.05 and 3.24 for liquid and solid states correspondingly) and $v_2$ (λ ~ 6.06 ÷ 6.27 in dependence on phase state).

\* \* \*

Now we can briefly consider some other gases that can be organized in clusters or even form drops in the atmosphere: $CO_2$, $CH_4$, $N_2O$, their possible density and temperatures in the upper atmosphere can be, in principle, sufficient for such transitions.

For $CO_2$, the carbon dioxide, $\Lambda^{(subl)}$ = 25.23 kJ/mole and by (3) $\lambda_1$ = 4.76 mcm, $\lambda_2$ = 2.38 mcm. Observable peaks are located between 4.3 ÷ 4.5 mcm, 2.4 ÷ 2.8 mcm. Thus a qualitative correspondence can be claimed. The difference can be connected with the energy needed for bend of pure linear molecule in solid substance.

For $N_2O$, the nitrous oxide, $\Lambda^{(subl)}$ = 23 kJ/mole, $\Lambda^{(cond)}$ = 16.56 kJ/mole and correspondingly $\lambda^{(s)}_1$ = 5.2 mcm and $\lambda^{(c)}_1$ = 7.25 mcm. They also correspond to observable peaks at 5 and 8 mcm.

For $CH_4$, the methane, $\Lambda^{(subl)}$ = 8.22 kJ/mole and by (3) $\lambda_1$ = 14.6 mcm, but observable peaks are located near to 8 and 5 mcm. Formally they can be attributed to condensation of dimers, trimers, and so on. But for more realistic consideration the knowledge of spectra of corresponding crystals are needed.

**CONCLUSIONS AND PERSPECTIVES**

Described results and certain further perspectives can be so summed.

1. All peaks of water vapors and of some air pollutions are identified as relating to phase transitions.

2. It can allow use of simple methods for detection of air pollutions and can be generalized by consideration of other gases.

3. As processes of solidification and evaporation can be reiterated and their repeating lead toward several mechanisms of reorganization or canalization of electromagnetic energy in the atmosphere. This conclusion requires re-examination of all heat balance and relative roles of different channels of heat removal that can have special interests at a vital discussion of climatic changes.

4. Considered processes do not exhaust all potentialities of electromagnetic radiation accompanying phase transitions. But others can be or even must be connected with sufficiently lower energy (cf. [10]).

5. For further development of offered theory some experiments seem needed. The most simple between them can be the establishing of difference of "gases peaks" during daylight hours and nightly. But laboratory investigation of latent energy emission should be also very desirable.

6. The offered theory can has direct significance for examination of planetary and nebulae atmospheres. So, it can be applied, in particular, to the very old problem of the Great Red Spot of Jupiter: its coloration can be connected, at least particularly, with processes of condensation and crystallization of ammonia and its compounds. But such approach requires laboratory investigations.

**Acknowledgement**

The author wish to thank Dr. Jeffrey Lapides for discussions these problems.